\documentclass[12pt]{JHEP3}
\usepackage{amsmath,amsfonts,amssymb}
\usepackage{verbatim,graphicx,float}  

\title{Universe from vacuum in loop-string cosmology}

\author{Jakub Mielczarek$^{a,b,\dagger}$ and Marek Szyd{\l}owski$^{a,c,d,\star}$\\ 
$^a$ Astronomical Observatory, Jagiellonian University, 30-244
Krak\'ow,\\ Orla 171, Poland \\
$^b$ Institute of Physics, Jagiellonian University, 30-059 
Krak\'ow,\\ Reymonta 4, Poland \\
$^c$ Department of Theoretical Physics, 
Catholic University of Lublin, \\
 Al. Rac{\l}awickie 14, 20-950 Lublin, Poland \\
$^d$ Marc Kac Complex Systems Research Centre, Jagiellonian University,\\
Reymonta 4, 30-059 Krak{\'o}w, Poland 
\\

$^{\dagger}$ \email{jakubm@poczta.onet.pl} \\
$^{\star}$ \email{uoszydlo@cyf-kr.edu.pl}
}

\abstract{
In this paper we study the description of the Universe based on the 
low energy superstring theory  modified by the Loop Quantum Gravity effects.
This approach was proposed by De Risi et al. in the Phys.\ Rev.\  D {\bf 76} 
(2007) 103531. We show that in the contrast with the 
string motivated pre-Big Bang scenario, the cosmological realisation of the 
$t$-duality transformation is not necessary to avoid an initial singularity. In 
the model considered the universe starts its evolution in the vacuum phase at 
time $t\rightarrow - \infty$. In this phase the scale factor $a\rightarrow 0$, 
energy density $\rho \rightarrow 0$ and coupling of the interactions 
$g^2_s \rightarrow 0$. After this stage the universe evolves to the non-singular 
hot Big Bang phase $\rho \rightarrow \rho_{\text{max}} < \infty$. Then the 
standard classical universe emerges. During the whole evolution the scale factor 
increases monotonically. We solve this model analytically. We also propose 
and solve numerically the model with an additional dilaton potential in which 
the universe starts the evolution from the asymptotically free vacuum phase 
$g^2_s \rightarrow 0$ and then evolves non-singularly to the emerging dark 
energy dominated phase with the saturated coupling constant 
$g^2_s \rightarrow \text{const}$.}

\begin{document}

\section{Introduction} \label{sec:Intro}

It is commonly believed and physically motivated that gravity exhibits quantum 
nature close to the Planck energy scale. However it can be realised that 
such an energy scale will be never reached with the present generation of 
accelerators. The fact that there is lack of experimental verification, is 
the main handicap in development of the quantum version of General Relativity. 
However according to the standard Big Bang scenario, our Universe arose from 
the epoch in which quantum gravitational description is adequate. This 
potentially allows for the verification of the quantum gravitational models by 
the astronomical observations. On the other side the application of the quantum 
gravity effects in the description of the universe could help to answer the 
most fundamental questions about the origin of the Universe. In particular one 
expects that the cosmological singularity should be avoided. These facts make 
investigation of the quantum gravitational models of the Universe so important 
and fascinating. 

Many efforts have been made to tackle this issue. One of the most promising 
propositions was the application of the Superstring Theory (ST). 
In this approach, called the pre-Big 
Bang cosmology \cite{Gasperini:2002bn,Gasperini:2007vw,Lidsey:1999mc,Gasperini:2007ar},
the $t$-duality has been applied leading to the avoidance of the initial 
singularity. The resulting dynamics possess a non-singular type of evolution with the 
minimal universe scales corresponding to the self-dual point of the $t$-duality 
transformation. Another interesting approach is based on Loop Quantum Gravity (LQG)
\cite{Ashtekar:2004eh} which is a promising candidate for the quantum theory of 
gravitation. The methods of LQG applied in the cosmological context lead to the 
avoidance of the initial singularity \cite{Bojowald:2006da}. In this scenario 
the universe is initially in the low energy contracting phase, reaches a 
minimal size and due to quantum effects evolves toward the expanding low energy 
phase \cite{Ashtekar:2006rx,Ashtekar:2006uz,Ashtekar:2006wn}. In the case 
with the cosmological constant this model was studied in \cite{Bentivegna:2008bg}.
It was shown that bouncing behaviour 
occurs for a dense subspace of the physical Hilbert space and semi-classicality 
is preserved across the bounce \cite{Ashtekar:2007em,Corichi:2007am}. The 
similar quantum bouncing behaviour is also recovered in Bojowald's approach,
for a recent review of this issue see \cite{Bojowald:2008pu,Bojowald:2008ma}.

In this paper we combine these two approaches to construct a non singular 
model of the universe without the need for the non-deterministic $t$-duality 
transformation. This idea, that the LQG effects can have important influence 
for the pre-Big Bang cosmology, was originally proposed by De Risi et al.
\cite{De Risi:2007gp} \footnote{Early works in LQC also indicate
singularity resolution for brane collision model \cite{Bojowald:2004kt}.}.

In the present paper we investigate the model of De Risi et al. in more detailed way 
and show some new resulting consequences. Especially we find the analytic 
solution of this model and study numerically effects of the dilaton potential. 

The presented approach is based on replacement of the starting point action. 
Namely the LQG is the background independent quantisation of the 
Hilbert-Einstein action
\begin{equation}
S[g_{\mu\nu}]= \frac{1}{2\kappa} \int d^4x \sqrt{-g} R.
\label{HEaction}
\end{equation}
On the other side, ST of the closed strings leads to the gravitational action 
in the form \cite{Lidsey:1999mc,Copeland:1994vi}
\begin{equation}
S[\phi,g_{\mu\nu},B_{\mu\nu}]= \frac{1}{2\kappa_D} \int d^Dx \sqrt{-g} e^{-\phi} \left[R+(\nabla \phi)^2-V(\phi)
-\frac{1}{12}H^2  \right]
\label{ESTaction}
\end{equation}
where $H^2=H_{\mu\nu\alpha}H^{\mu\nu\alpha}$ and $H_{\mu\nu\alpha}=3! \partial_{ [ \mu} B_{\nu\alpha ] }$.
Here beside the metric field $g_{\mu\nu}$ two additional fields occur. 
Namely the dilaton scalar field $\phi$ and the asymmetric Kalb-Ramond field $B_{\mu\nu}$.
In case when these fields vanish action (\ref{ESTaction}) take a form of the 
action (\ref{HEaction}). Otherwise the scalar field is non-minimally coupled 
with the curvature and can lead to the important modifications of dynamics. 
It is worth to note that action (\ref{ESTaction}) is related 
with the Brans-Dicke theory with the B-D parameter $\omega=-1$. This means 
that it can be considered as a theory with the dynamical coupling constant 
defined as
\begin{equation}
g^2_s = e^{\phi}.
\label{coupling}
\end{equation}     
In the case when $\phi \rightarrow -\infty$ then $g^2_s  \rightarrow 0$ and 
asymptotic freedom occurs. On the other side $\phi \rightarrow \infty$ leads 
to the confinement. In the present Universe we expect $g^2_s \simeq$ const.

In the presented approach we consider action (\ref{ESTaction}) instead of (\ref{HEaction}) 
as a starting point for the loop quantisation. This means that we can consider 
ST as a fundamental theory and the LQG as a background independent method of the 
quantisation of the metric field. However this interpretation is not unique.
In fact we do not even need to assume ST to be valid but rather to postulate
form of the action. In this paper we restrict our preliminary investigations 
to the graviton-dilaton sector in the $D=4$. However the Kalb-Ramond field is 
not coupled non-minimally to the gravitational field and has less important 
influence for dynamics.   

The organisation of the text is the following. In section \ref{sec:loop} we 
introduce holonomy corrections to the action (\ref{ESTaction}) in the Einstein 
frame. Then in section \ref{sec:vacuum} we investigate analytically resulting
cosmological model. In section \ref{sec:darkenergy} we improve the analytical
model by introduction of the additional potential for the dilaton field.
In section \ref{sec:summary} we summarise the results.

\section{String cosmology with holonomy corrections} \label{sec:loop}

In this section we show how to introduce phenomenological effects of the 
quantum holonomies in the classical equations of motions. We also give the 
heuristic explanation of the form of these corrections. In result we obtain the 
phenomenological Hamiltonian which contains information about the effective 
form of the quantum gravitational effects. The resulting dynamics trace the 
mean value of the quantum state. However to obtain information about 
dispersions and approve semi-classical considerations full quantum treatment 
should be performed.  

To introduce loop modifications to the action (\ref{ESTaction}) we choose the strategy
already applied to the theory with non-minimally coupled scalar fields 
\cite{Bojowald:2006bz,Bojowald:2006hd}. The presented approach is also equivalent 
to that presented in the paper \cite{De Risi:2007gp}. 
Namely, we perform the conformal transformation in the form
\begin{equation}
\tilde{g}_{\mu\nu}  = \exp \left(-\frac{2\phi}{D-2}\right) g_{\mu\nu}
\end{equation}
which transforms the effective string action (\ref{ESTaction}) from so called 
string frame to the Einstein frame where
\begin{equation}
S= \frac{1}{2\kappa_D} \int d^Dx \sqrt{-\tilde{g}}\left[ \tilde{R}-\frac{1}{D-2}(\tilde{\nabla} \phi)^2
-V(\phi)  e^{\frac{2\phi}{D-2}}-\frac{1}{12}\tilde{H}^2 e^{-\frac{4\phi}{D-2}}  \right].
\end{equation}
We restrict now our preliminary consideration to the graviton-dilaton sector 
in the $D=4$ case, then
\begin{equation}
S= \int dt L = 
 \frac{1}{2\kappa_4} \int d^4x \sqrt{-\tilde{g}}\left[ \tilde{R}-\frac{1}{2}(\tilde{\nabla} \phi)^2
-V(\phi) e^{ \phi }  \right].
\end{equation}
In this case $\kappa_4 g^2_s=\kappa=8\pi G$ where $G$ is the classical Newton coupling 
constant. To quantise this theory in the background independent way we perform 
the Legendre transformation of the L to the canonical formulation. Then we 
express the obtained Hamiltonian in terms of the Ashtekar variables $(A,E)$ \cite{Ashtekar:1987gu}
which take value in $\mathfrak{su}(2)$ and $\mathfrak{su}(2)^*$ algebras respectively. 
The full Hamiltonian is the sum of the gravitational part $\mathcal{H}_{\text{G}}$ 
and dilaton part $\mathcal{H}_{\phi}$. In terms of the Ashtekar variables the 
Hamiltonian for general relativity is a sum of constraints
\begin{equation}
\mathcal{H}_{\text{G}} = \int d^3 {\bf x} \, (N^i G_i + N^a C_a + N  h_{\text{sc}}),
\end{equation}
where 
\begin{align}
C_a &= E^b_i F^i_{ab} - (1-\gamma^2)K^i_a G_i ,\nonumber \\
G_i &= D_a E^a_i
\end{align}
and the scalar constraint has a form
\begin{align}\label{ham}
&\mathcal{H}_{\text{S}}:=\int d^3{\bf x} \, N(x) h_{\rm sc}= \nonumber \\ 
  &\frac{1}{2\kappa_4} \int d^3{\bf x} \, N(x)\left( \frac{E^a_i
  E^b_j}{\sqrt{|\det E|}} {\varepsilon^{ij}}_k F_{ab}^k -
  2(1+\gamma^2) \frac{E^a_i E^b_j}{\sqrt{|\det E|}} K^i_{[a}
  K^j_{b]} \right) 
\end{align}
with the curvature of the Ashtekar connection $F=dA + \frac{1}{2}[A,A]$ and where 
$\gamma$ is the Barbero-Immirzi parameter. The dilaton Hamiltonian has the form
\begin{equation}
\mathcal{H}_{\tilde{\phi}}=\int d^3{\bf x} \, N(x)\left( \frac{1}{2}\frac{\pi^2_{\tilde{\phi}}}{\sqrt{|\det E|}}  +
 \frac{1}{2} \frac{E^a_i E^b_i \partial_a \tilde{\phi} \partial_b \tilde{\phi}  }{\sqrt{|\det E|}}
  + \sqrt{|\det E|} U(\tilde{\phi}) \right)
\end{equation}
where we have defined
\begin{eqnarray}
\tilde{\phi}   &=&  \frac{\phi}{\sqrt{2\kappa_4}}, \\
U(\tilde{\phi}) &=&  \frac{1}{2\kappa_4} V(\phi)e^{\phi}.
\end{eqnarray}

The next step is to define a spacetime symmetry.
It is worth to note that if we assume a given symmetry in the string frame 
it will be the same in the Einstein frame. The only difference is the transformation
of time $t \rightarrow \tilde{t}$ and the scale factor $a \rightarrow \tilde{a}$ for 
which the form of the metric holds.  
We choose the flat FRW $k=0$ spacetime for which metric can be written as
\begin{equation}
d\tilde{s}^2=-N^2(x) d\tilde{t}^2 + q_{ab}dx^adx^b
\end{equation}
where $N(x)$ is the lapse function and the spatial part of the metric is expressed as 
\begin{equation}
q_{ab}= \delta_{ij} {\omega^i_a} {\omega^j_b}= \tilde{a}^2(t) {^oq}_{ab} =
 \tilde{a}^2(t)  \delta_{ij}  {^o\omega^i_a}{^o\omega^j_b}.
\end{equation}
In this expression ${^oq}_{ab}$ is the fiducial metric and ${^o\omega^i_a}$ are the co-triads dual to the triads 
${^oe^a_i}$,  ${^o\omega^i}({^oe_j})=\delta^i_j$   where $^o\omega^i={^o\omega^i_a}dx^a$ and $^oe_i={^oe_i^a}\partial_a$.
In this case the Ashtekar variables take the form
\begin{eqnarray}
A &\equiv& \Gamma +\gamma K  = {c} V_0^{-1/3} \ {^o\omega^i_a}  \tau_i dx^a , \label{A} \\
E &\equiv& \sqrt{|\det q|} e  = {p} V_0^{-2/3} \sqrt{^oq} \ {^oe^a_i} \tau_i  \partial_a \label{E}
\end{eqnarray}
where $V_0$ is the volume of fiducial cell.
The volume $V_0$ is just a scaling factor and can be chosen
arbitrary in the domain $V_0\in \mathbb{R}_+$.
The physical results do not depend on the choice of $V_0$.
The pair $(c,p)$ are canonical 
variables for the gravitational field and can be expressed
in terms of the standard FRW variables  $(c,|p|)=(\gamma \dot{\tilde{a}} V^{1/3}_0,\tilde{a}^2V^{2/3}_0 )$.
Now it is straightforward to calculate
the full classical Hamiltonian in the canonical variables $(c,p,\tilde{\phi},\tilde{\pi})$ 
\begin{equation}
\mathcal{H} = -\frac{3}{\kappa_4 \gamma^2} \sqrt{|p|}c^2 +\frac{{\pi}^2_{\tilde{\phi}}}{|p|^{3/2}} +|p|^{3/2}U(\tilde{\phi}) 
\end{equation}
where we assumed homogeneity of the field $\tilde{\phi}$ and chosen gauge 
$N(x)=1$. Loop quantisation of such a model in case of the free field was performed 
in the works \cite{Ashtekar:2006rx,Ashtekar:2006uz,Ashtekar:2006wn}. Here we 
only sketch the main steps and for the detailed considerations we send the reader 
to the mentioned papers. To quantise this theory in the background independent 
way one introduces holonomies of connection $A$ 
\begin{equation}
h_{\alpha}[A] = \mathcal{P} \exp \int_{\alpha} A  \ \ \text{where 1-form} \ \   A=\tau_i A^i_a dx^a
\label{holo}
\end{equation}
and conjugated fluxes     
\begin{equation}
F_{S}^i[E] = \int_S dF^i   \ \ \text{where 2-form} \ \  dF_i =\epsilon_{abc} E^a_i dx^b \wedge dx^c 
\end{equation}
where $\alpha,S \in \Sigma$ and $2 i \tau_i= \sigma_i$ where $\sigma_i$ are the Pauli matrices.
From this definition we can calculate holonomy in the direction $^oe^a_i\partial_a$ and the length $\mu V_0^{1/3}$ 
\begin{eqnarray}
h_{i}^{(\mu)} = \exp \int_0^{\mu V_0^{1/3} } \tau_i c V_0^{-1/3}  {^o\omega^i_a} dx^a 
              = \mathbb{I}\cos \left( \frac{\mu c}{2}\right)+2\tau_i\sin \left( \frac{\mu c}{2}\right)
\label{hol2}
 \end{eqnarray}
where we used the definition of the Ashtekar variable $A$ (\ref{A}). From such 
a particular holonomies we can construct holonomy along the closed curve 
$\alpha=\Box_{ij}$. This holonomy can be written as 
\begin{eqnarray}
h_{\Box_{ij}}^{(\mu)} = h_{i}^{(\mu)} h_{j}^{(\mu)} h_{i}^{(\mu)-1} h_{j}^{(\mu)-1} .
\end{eqnarray}

Now it is straightforward to show that the field strength can be expressed as 
\begin{equation}
F^k_{ab} = - 2 \lim_{Ar \rightarrow 0}
 \frac{\text{tr}\left[\tau_k \left( h^{(\mu)}_{\Box_{ij}}-\mathbb{I} \right) \right]}{\mu^2 V_0^{2/3} } 
{^o\omega^i_a}{^o\omega^j_b}.
\label{lim}
\end{equation}
The trace in this equation can be explicitly calculated  
\begin{equation}
\text{tr}\left[\tau_k \left( h^{(\mu)}_{\Box_{ij}}-\mathbb{I} \right)  \right] =
 - \frac{\epsilon_{kij}}{2} \sin^2\left(\mu c \right).
\label{tr}
\end{equation}
In Loop Quantum Cosmology the limit $Ar \rightarrow 0$ in the formula (\ref{lim}) 
does not exist because of existence of the area gap. The area gap corresponds 
to the minimal quanta of area $\Delta=2\sqrt{3}\pi\gamma l_{\text{Pl}}^2$ \cite{Ashtekar:1996eg}.
So instead of the limit in equation (\ref{lim}) we should stop shrinking the
loop at the appropriate minimal area $\Delta$. This area corresponds to the 
area $\boxdot_i$ intersected by the loop and to take account discreetness of the
space we should perform a limit $\boxdot_i \rightarrow \Delta$.
Now, we must connect the area $\boxdot_i$ with the length $\mu$ of the loop edge. 
We can choose that area $\boxdot_i$ correspond to the physical area $\tilde{a}^2 \mu^2$. 
So in this case $\boxdot_i=V_0^{2/3} \tilde{a}^2 \mu^2 = |p|\mu^2$ we have in the limit  
\begin{equation}
\mu = \bar{\mu}  := \sqrt{ \frac{\Delta}{|p|}}.
\end{equation}
We will use this value in the further consideration.
However we must to mention that the choice of the function $\bar{\mu}$ is not unique and 
leads to ambiguities. However function $\bar{\mu}$ used here leads 
to the proper classical limit and follows from loop quantization \cite{Ashtekar:2006wn}.
Special situation corresponding to so called $\mu_{o}-$scheme where $\mu=\mu_{o}=$const.
In such a case $\mu_0$ depend explicitly on the factor $V_0$.
However it was shown that $\mu_{o}-$scheme does not have the 
classical limit and cannot be treated physically \cite{Ashtekar:2006uz,Mielczarek:2008zv}.  
In general ambiguities of $\mu(p)$ come from consideration of the so called 
lattice models \cite{Bojowald:2006qu} for which in general 
$\mu(p) \propto |p|^{n}$ where $n \in [-1/2,0]$ \cite{Bojowald:2008pu}. 
However recent results indicate that the only $\bar{\mu}$ approach is consistent
quantization scheme \cite{Corichi:2008zb}.
The issue of other kind of ambiguities was addressed in the papers
\cite{Mielczarek:2008mf,Hrycyna:2008yu}. 
 
Based on the above considerations we conclude that phenomenological effects of 
the quantum holonomies can be introduced by a replacement
\begin{equation}
c \rightarrow  \frac{ \sin \left( \bar{\mu} c  \right) }{\bar{\mu}} 
\label{holcor}
\end{equation}
in the classical expression. In fact another type of corrections occur
in the Loop Quantum Cosmology. However there is no corrections in the flat models.
Namely the inverse volume corrections which 
should be in general included have shown to be physically inconsistent 
in the context of the flat models. This is the result of the non-compact 
and flat manifold $\Sigma$. In the cosmological models with the internal 
curvature like FRW $k=\pm 1$ inverse volume corrections should be also 
taken into account \cite{Ashtekar:2006es}.    
 
\section{Non-singular universe from vacuum} \label{sec:vacuum}

\begin{figure}[ht!]
\centering
\includegraphics[width=10cm,angle=0]{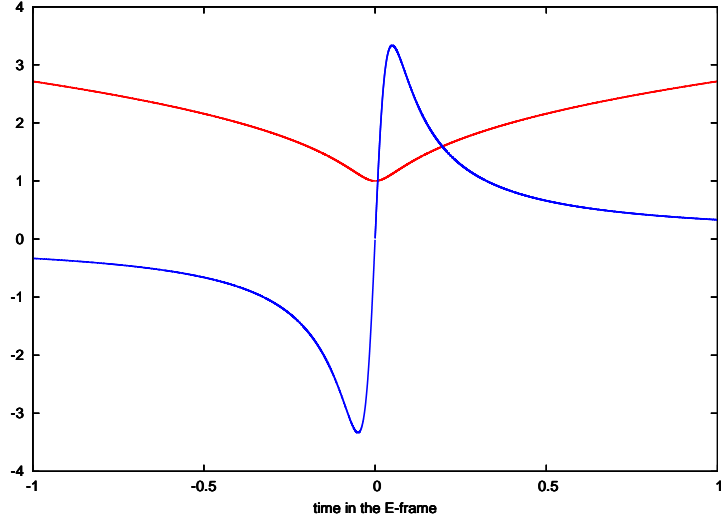}
\caption{The top curve (red) presents the bouncing solution for the $\tilde{a}(\tilde{t})$.
Bottom curve (blue) corresponds to the time $\tilde{t}$ evolution of the Hubble parameter 
$\tilde{H}=\frac{1}{\tilde{a}}\frac{d\tilde{a}}{d\tilde{t}}$.}
\label{fig:1}
\end{figure}

The introduction of the holonomy corrections investigated in the previous 
section lead to the effective Hamiltonian in the form
\begin{equation}
\mathcal{H}_{\text{eff}} =  - \frac{3}{\kappa_4 \gamma^2} \sqrt{|p|} \left[ \frac{ \sin \left( \bar{\mu} c
\right) }{\bar{\mu}}   \right]^2 
+ \frac{1}{2} \frac{ \pi_{\tilde{\phi}}^2}{  {|p|}^{3/2} }
\label{model}
\end{equation}
where we assumed  homogeneity of the dilaton field.
This Hamiltonian fulfils constraint $\mathcal{H}_{\text{eff}} = 0$. 
The equations of motion can be now derived with use of the Hamiltonian equation 
$\dot{f}=\left\{f , \mathcal{H}_{\text{eff}} \right\}$ and for 
the canonical variables are
\begin{eqnarray}
\dot{p} &=& \left\{p  , \mathcal{H}_{\text{eff}} \right\} = 
 \frac{2}{\gamma} \frac{ \sqrt{|p|} }{\bar{\mu} }  \sin \left( \bar{\mu} c 
\right) \cos \left( \bar{\mu} c  \right), \\
\dot{c} &=& \left\{c  , \mathcal{H}_{\text{eff}} \right\}= 
 - \frac{1}{\gamma} \frac{\partial}{\partial p} \left\{ \sqrt{|p|} 
\left[ \frac{ \sin \left( \bar{\mu} c \right)  }{\bar{\mu}}   \right]^2   \right\}
-\text{sgn}(p)\frac{\kappa_4 \gamma}{4}  \frac{\pi^2_{\tilde{\phi}}}{{|p|}^{5/2}} ,  \\
\dot{\tilde{\phi}} &=& \left\{ \tilde{\phi}  , \mathcal{H}_{\text{eff}} \right\}   =
   {|p|}^{-3/2} \pi_{\tilde{\phi}},  \\
\dot{\pi}_{\tilde{\phi}} &=& \left\{ \pi_{\tilde{\phi}} , \mathcal{H}_{\text{eff}} \right\} = 0 
\label{equations}
\end{eqnarray}
where all derivatives are in respect to the time $\tilde{t}$. The exact solution for 
the variable $p$ can be find \cite{Mielczarek:2008zv} and has a form 
\begin{eqnarray}
p(\tilde{t}) &=& \text{sgn}(p)   \left[ \frac{1}{6}\kappa_4\pi_{\tilde{\phi}}^2\gamma^2 \Delta +
\frac{3}{2} \kappa_4 \pi_{\tilde{\phi}}^2 \tilde{t}^2  \right]^{1/3}.
\label{sol1}
\end{eqnarray}
This corresponds to the bouncing type behaviour in the Einstein frame. 
In Fig. \ref{fig:1} we show evolution in this frame for the scale factor 
$\tilde{a}$ and the Hubble parameter 
$\tilde{H}=\frac{1}{\tilde{a}}\frac{d\tilde{a}}{d\tilde{t}}$.

\begin{figure}[ht!]
\centering
\includegraphics[width=10cm,angle=0]{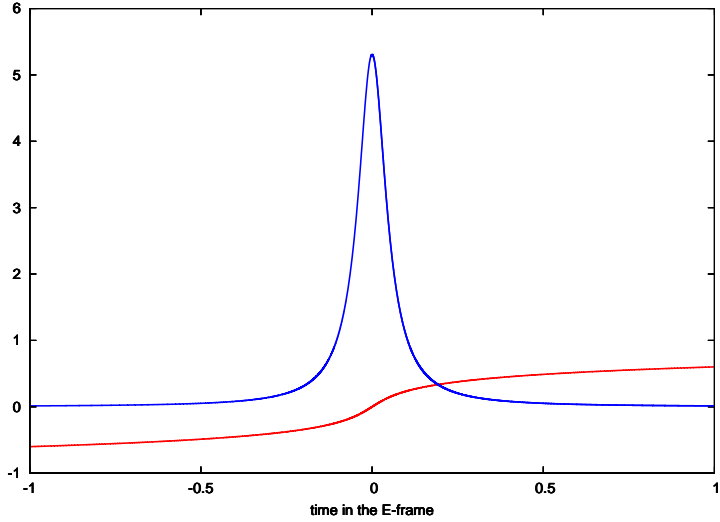}
\caption{Monotonic (red) curve presents the evolution of the
 dilaton field $\tilde{\phi}(\tilde{t})$.
Picked  (blue) curve shows the time $\tilde{t}$  dependence of the dilaton
 energy density $\rho = \dot{\tilde{\phi}}^2/2$.}
\label{fig:2}
\end{figure}

The scale factors and times in the Einstein and string frames are related by 
\begin{eqnarray}
 a &=& e^{\phi/2} \tilde{a},  \\
dt &=& e^{\phi/2} d \tilde{t}. \label{timetrans}
\end{eqnarray}
Therefore to analyse the evolution in the string frame the solution for the 
dilaton field is required. This can be obtained analytically by the simple 
integration
\begin{eqnarray}
\phi &=& \sqrt{2\kappa_4} \tilde{\phi}  =  \sqrt{2\kappa_4} \int d\tilde{t} \frac{\pi_{\tilde{\phi}}}{ {|p|}^{3/2} } \nonumber \\
 &=& \sqrt{2\kappa_4} \frac{\pi_{\tilde{\phi}}  }{\sqrt{ \frac{3}{2} \kappa_4 \pi_{\tilde{\phi}}^2 }  }
\int \frac{d\tilde{t}}{\sqrt{ \frac{1}{9} \gamma^2 \Delta + \tilde{t}^2  }} 
= \frac{2}{\sqrt{3}} \text{sgn}(\pi_{\tilde{\phi}}  ) \text{arcsh}
 \left( \frac{3 \tilde{t}}{\gamma \sqrt{\Delta} }  \right).
\end{eqnarray}
The solution depends on $\text{sgn}(\pi_{\tilde{\phi}}  )$ leading to monotonically increasing
or decreasing functions. In the case $\pi_{\tilde{\phi}}<0$ there is no birth of the classical universe.
Therefore in the further considerations we choose $\pi_{\tilde{\phi}}>0$. In this case $\phi(\tilde{t})$ is 
a monotonically increasing function. We show this behaviour together with evolution of the 
energy density 
\begin{equation}
\rho = \frac{\dot{\tilde{\phi}}^2}{2}
\end{equation}
in Fig. \ref{fig:2}. The important feature of this evolution is that 
energy density does not diverge. The maximal value of the energy density 
\begin{equation}
\rho_{\text{c}} = \frac{3}{\kappa_4 \gamma^2 \Delta}
\end{equation}
corresponds 
to the hot Big Bang which is here the transitional non-singular phase. 
This result is with agreement with the results of the paper \cite{De Risi:2007gp}. 

The subsequent quantity that we can investigate is time $\tilde{t}$ dependence of the 
coupling constant (\ref{coupling}) which is given by
\begin{equation}
g^2_s = \exp \left\{ \frac{2}{\sqrt{3}} \text{sgn}(\pi_{\tilde{\phi}}  )  \text{arcsh}
 \left( \frac{3 \tilde{t}}{\gamma \sqrt{\Delta} }  \right)   \right\}. 
\end{equation} 
This function rapidly decreases to zero for negative value of time $\tilde{t}$
leading to the asymptotic lack of the interactions in the pre-Big Bang phase.
In the post Big Bang phase the coupling constant $g^2_s$ grows monotonically.
This is rather non required behaviour, because we rather expect 
the coupling constant to saturate. In the next section we show 
that when the dilaton potential is present such a scenario can occur.
The presented here model can be rather treated as an analytical toy model
which requires some further modifications and analysis. 
 
\begin{figure}[ht!]
\centering
\includegraphics[width=10cm,angle=0]{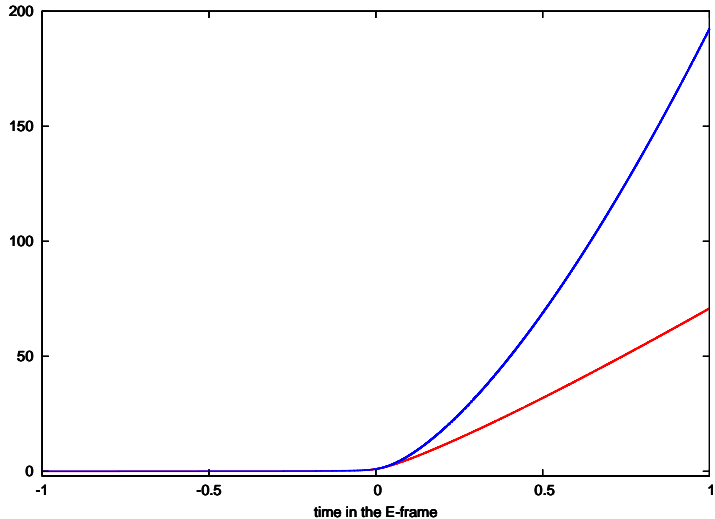}
\caption{The top curve (blue) shows the evolution of the scale factor $a(\tilde{t})$.
The bottom curve (red) shows the time $\tilde{t}$ dependence of the coupling constant $g^2_s$.}
\label{fig:3}
\end{figure}

The evolution of the scale factor in the string frame takes a form
\begin{eqnarray}
a = V^{-1/3}_0  \exp \left\{ \frac{\text{sgn}(\pi_{\tilde{\phi}}  )  }{\sqrt{3}} \text{arcsh}
 \left( \frac{3 \tilde{t}}{\gamma \sqrt{\Delta} }  \right)   \right\}  
 \left[ \frac{1}{6}\kappa_4\pi_{\tilde{\phi}}^2\gamma^2 \Delta +
\frac{3}{2} \kappa_4 \pi_{\tilde{\phi}}^2 \tilde{t}^2  \right]^{1/6}.
\label{evola}
\end{eqnarray}
We show this behaviour together with $g^2_s(\tilde{t})$ in Fig. \ref{fig:3}.
In the early times $\tilde{t} \rightarrow  -\infty$ the scale factor $a(\tilde{t} )$
goes to zero. This state can be interpreted as a vacuum state without interaction and
with vanishing energy density. It is important to note that this is the classical state
and the quantum effects are important only in the neighbourhood of the Big Bang
transition state. 

All above considerations we have performed in the time $\tilde{t}$. However we can 
back to the time $t$ by integration
 \begin{eqnarray}
t &=& \int d\tilde{t}  \exp \left\{ \frac{\text{sgn}(\pi_{\tilde{\phi}}  )  }{\sqrt{3}} \text{arcsh}
 \left( \frac{3 \tilde{t}}{\gamma \sqrt{\Delta} }  \right)   \right\}  \nonumber  \\
  &=& \frac{1}{2} \left[ 3 \tilde{t} 
-\frac{\text{sgn}(\pi_{\tilde{\phi}}  )  }{\sqrt{3}} \sqrt{ \gamma^2\Delta +9 \tilde{t}^2  } 
  \right] \exp \left\{ \frac{\text{sgn}(\pi_{\tilde{\phi}}  )  }{\sqrt{3}}
 \text{arcsh}\left( \frac{3 \tilde{t}}{\gamma \sqrt{\Delta} }  \right)   \right\}. 
\label{evoltime}
\end{eqnarray}
This is the monotonic function of $\tilde{t}$ so the results obtained with
time $\tilde{t}$ are equivalent with these in time $t$. Considering asymptotic 
behaviour of the formula (\ref{evola}) and (\ref{evoltime}) we recover 
evolution of the post-Big Bang Universe 
\begin{equation}
a(t) \propto t^{\frac{1}{\sqrt{3}}}  \approx t^{0.577}
\end{equation}
for $t \gg 0$. The asymptotic behaviour of the dilaton has a form
\begin{equation}
\phi(t)  \propto \ln \left(t \right).
\label{assymptphi}
\end{equation}
In the paper \cite{De Risi:2007gp} it was suggested that thanks to LQG effects, 
the dilaton in the string frame can approach a constant value for late times.
However solution (\ref{assymptphi}) indicates that such a scenario is not realised
in this simple model. The increase of dilaton field for the late times is however logarithmically
slow. For the late times the coupling constant behave as  
\begin{equation}
g^2_s  \propto t^{\sqrt{3}-1} \approx   t^{0.732}.  
\end{equation}     

In the next section we show how to improve this model to obtain the saturation 
of the coupling constant for the late times. 

\section{Possible evolutionary scenario: From asymptotically free to dark energy dominated universe} 
\label{sec:darkenergy}

In the previous section we have considered the model with the free dilaton field. 
Resulting dynamics was described by a simple analytical solution. To generalise 
this model it is naturally to add a non-vanishing potential function. This 
makes dynamics more complicated to analysis and requires numerical methods to 
be used. In this section we restrict ourselves to study qualitative 
modifications in dynamics that can occur when dilaton potential is present.

\begin{figure}[ht!]
\centering
\includegraphics[width=10cm,angle=0]{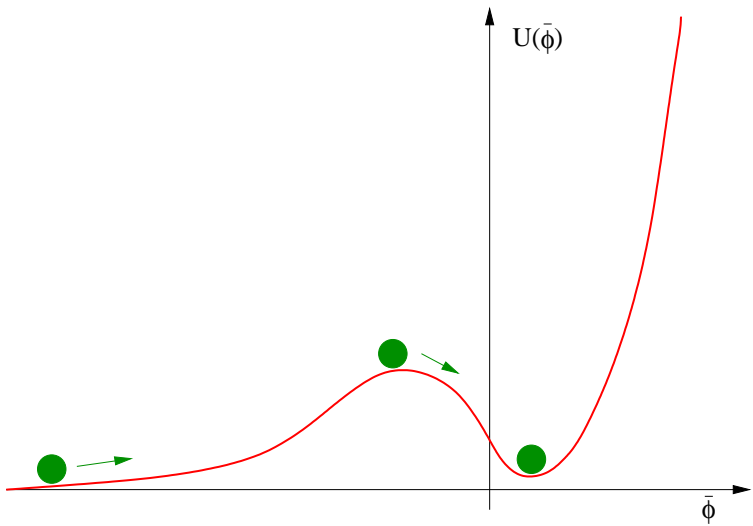}
\caption{Potential function $U(\tilde{\phi})$ and considered evolutionary scenario for the dilaton field.}
\label{fig:4}
\end{figure}

The simplest form of the potential energy that we can assume is the mass potential
\begin{equation}
V(\phi) = \frac{m^2}{2} (\phi-\phi_0)^2+V_0
\label{pot1}
\end{equation}
or the $\phi^4$ type interacting potential. In fact these two possibilities 
lead to the same qualitative prediction and we can concentrate on the 
potential (\ref{pot1}). In this case potential $U(\tilde{\phi})$ takes a form
\begin{equation}
U(\tilde{\phi})  = \frac{m^2}{2} (\tilde{\phi}-\tilde{\phi}_0)^2 e^{\sqrt{2\kappa_4}\tilde{\phi}}+
 \tilde{V}_0 e^{\sqrt{2\kappa_4}\tilde{\phi}}.
\label{pot2}
\end{equation} 
This potential together with the possible evolution path of the field is sketched 
in Fig. \ref{fig:4}.
The dynamics of the field $\tilde{\phi}$ is governed by the equation 
 \begin{equation}
\ddot{\tilde{\phi}}+3 \tilde{H}\dot{\tilde{\phi}}+\frac{dU(\tilde{\phi})}{d\tilde{\phi}}=0.
\label{phitildmot}
\end{equation}
This equation can in general contain quantum corrections form the inverse volume operator.
However they are physically inconsistent in the case of the flat models and 
are not expected to give contribution in this peculiar case.   
The dynamics of the dilaton filed is therefore governed by the classical formula.

The potential $U(\tilde{\phi})$ has two local minima, for
$\tilde{\phi}=\tilde{\phi}_0$ and for  $\tilde{\phi} \rightarrow  -\infty$. 
Lets imagine scenario in which field starts its evolution with asymptotically 
free state $\tilde{\phi} \rightarrow  -\infty$, $g^2_s\rightarrow 0$. In this phase 
the scale factor $a \rightarrow 0 $ and energy density $\rho \rightarrow 0$.
We can call this state asymptotically free vacuum. In this phase there is no 
interactions between particles and zero energy density in agreement with the 
hypothesis of ``asymptotic past triviality'' \cite{Buonanno:1998bi}. In the same 
time in the Einstein frame the scale factor decreases leading to the negative 
value of the Hubble parameter $\tilde{H}$. This leads to the negative friction 
term in equation (\ref{phitildmot}) resulting motion of the field toward higher 
values. The dynamics of the scalar field can be then approximated by 
$\ddot{\tilde{\phi}}+3 \tilde{H}\dot{\tilde{\phi}} \approx 0$.
Then energy density increases and the dilaton field reaches the top of the 
potential hill and subsequently falls down to the potential minima. 
This stage corresponds to the hot Big Bang phase
when energy density reaches its maximum. However there is no singular 
behaviour here. When the dilaton field falls to the the potential well
it dumps because $\tilde{H}$ starts to be positive.
It means that the non-singular hot Big Bang corresponds to the bounce in the
Einstein frame.

Let us investigate now the case with $V_0=0$ and $\phi_0=0$. The local maxima
for such a potential is placed in 
\begin{equation}
\tilde{\phi}_{\text{max}} =  - \sqrt{\frac{2}{\kappa_4}}
\end{equation}
what gives
\begin{equation}
U(\tilde{\phi}_{\text{max}}) =  \frac{m^2}{\kappa_4} \frac{1}{e^2}. 
\end{equation}
Because for physical evolution $\rho \leq \rho_{\text{c}}$, we obtain relation 
\begin{equation}
m^2 \leq \frac{\sqrt{3}e^2}{2\pi \gamma^3 G_4}.
\end{equation}
where $\kappa_4=8\pi G_4$.
This relation must be fulfilled to permit the transition of the field 
$\tilde{\phi}$ to the potential well around $\tilde{\phi}=0$.
In particular case $\gamma_{M}=0.12738$ \cite{Meissner:2004ju} it gives
\begin{equation}
m^2 \leq \frac{985,5}{G_4}.
\end{equation}
 
\begin{figure}[ht!]
\centering
$\begin{array}{cc}   
\includegraphics[width=7cm,angle=0]{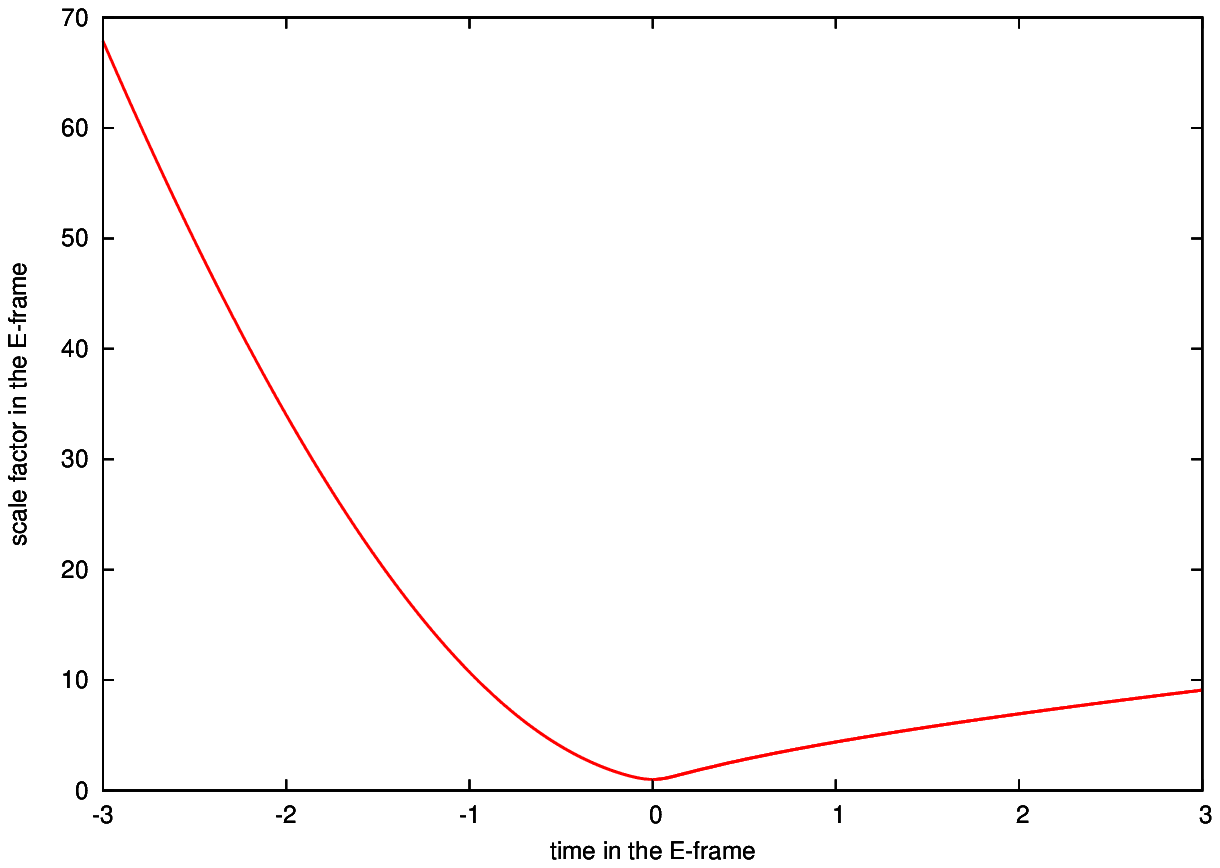}  &  \includegraphics[width=7cm,angle=0]{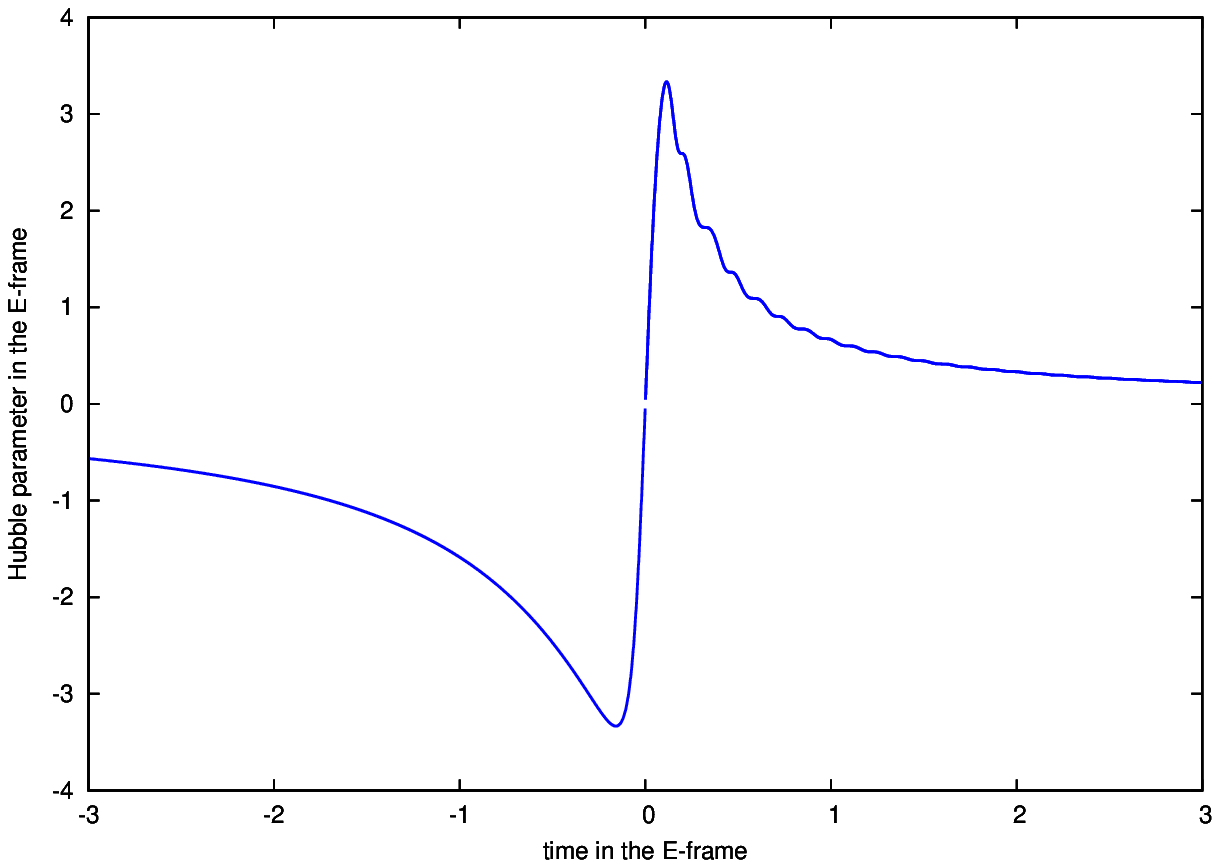}   
\end{array}
$
\caption{ 
{\bf Left  }: Evolution of the scale factor in the Einstein frame. 
{\bf Right }: Evolution of the Hubble parameter in the Einstein frame .}
\label{Numeric1}
\end{figure}

\begin{figure}[ht!]
\centering
$\begin{array}{cc}   
\includegraphics[width=7cm,angle=0]{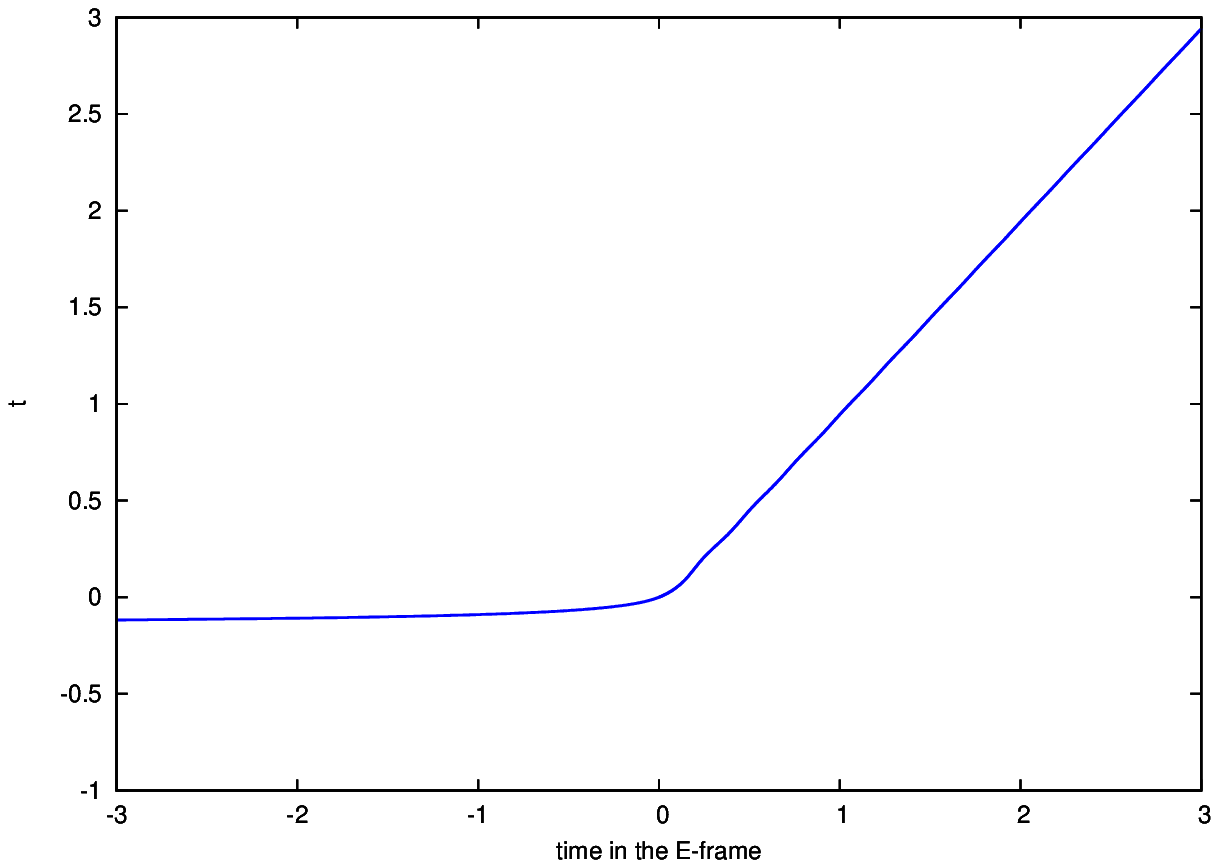}  &  \includegraphics[width=7cm,angle=0]{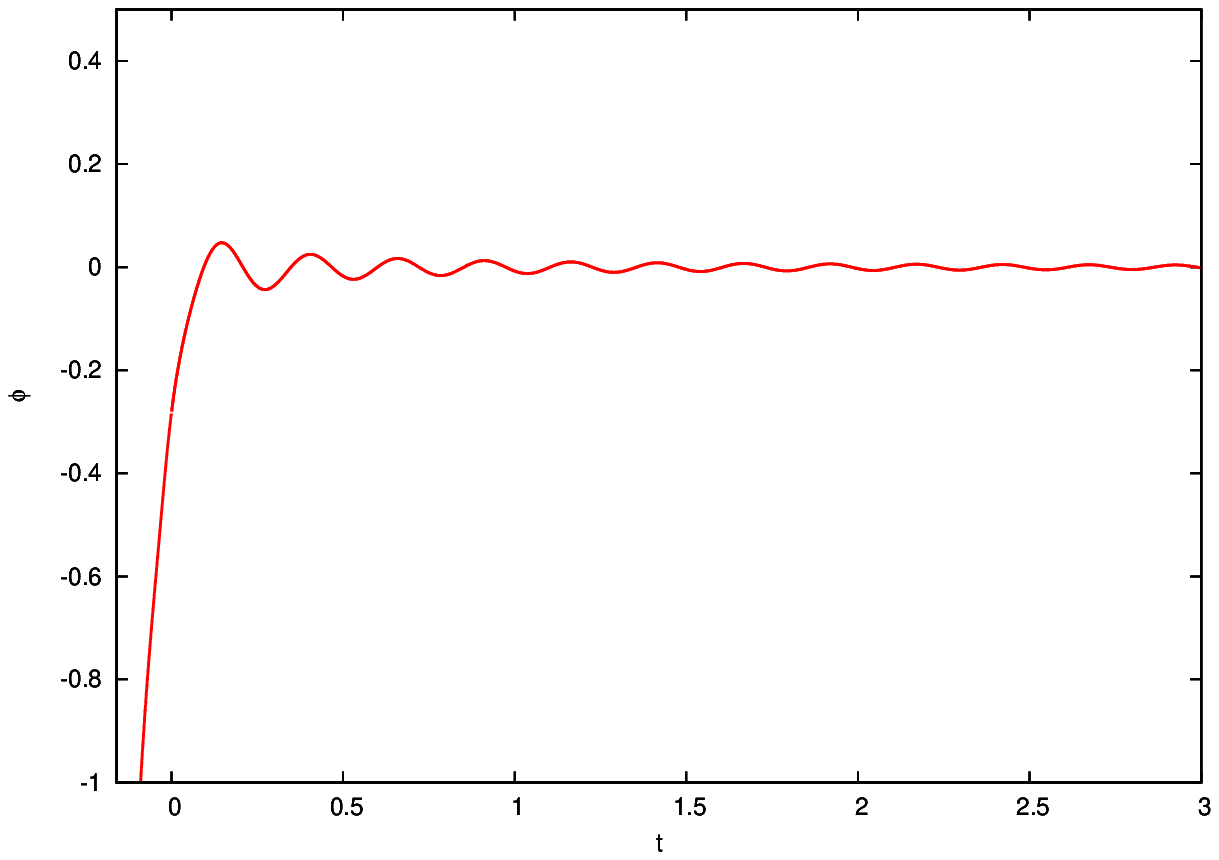}   
\end{array}
$
\caption{ 
{\bf Left  }: Dependence between time in the S-frame and time in the E-frame. 
{\bf Right }: Evolution of the field  $\phi$ in the time $t$.}
\label{Numeric2}
\end{figure}

\begin{figure}[ht!]
\centering
$\begin{array}{cc}   
\includegraphics[width=7cm,angle=0]{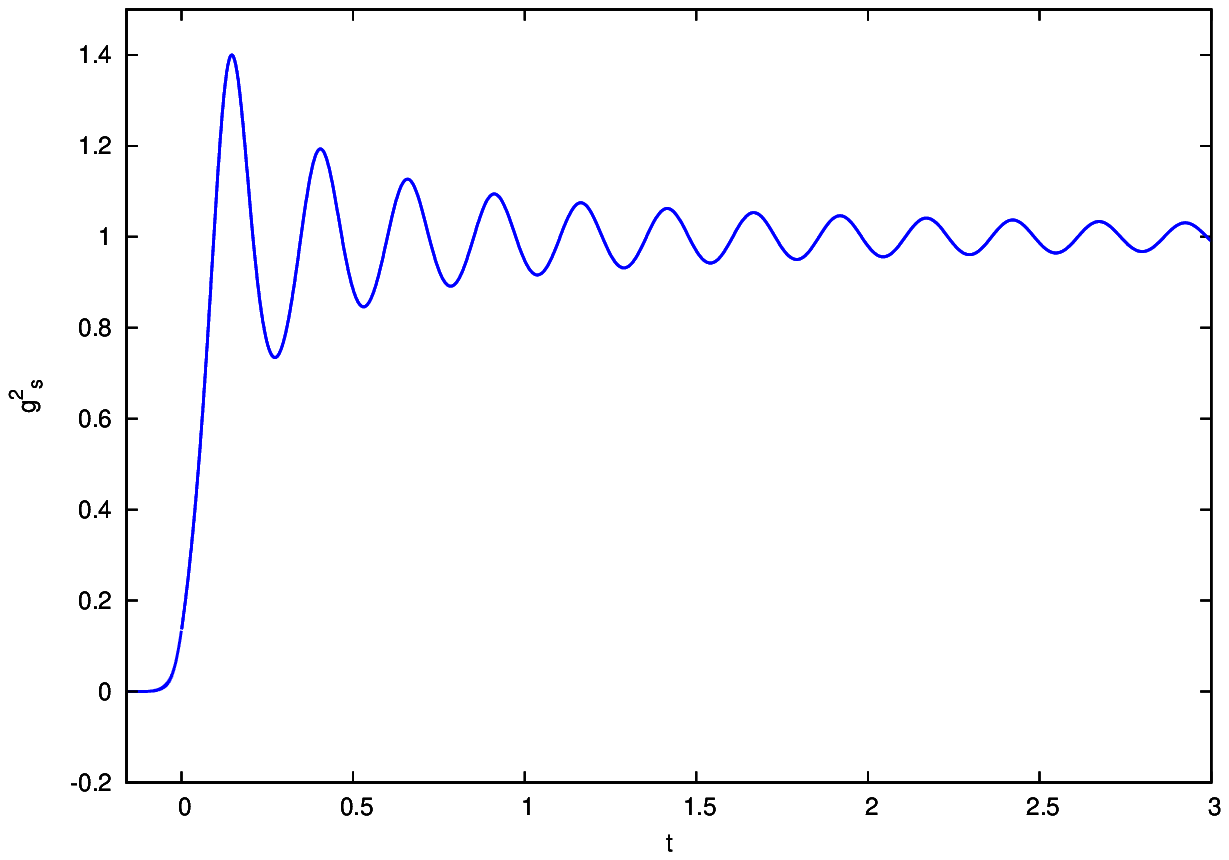}  &  \includegraphics[width=7cm,angle=0]{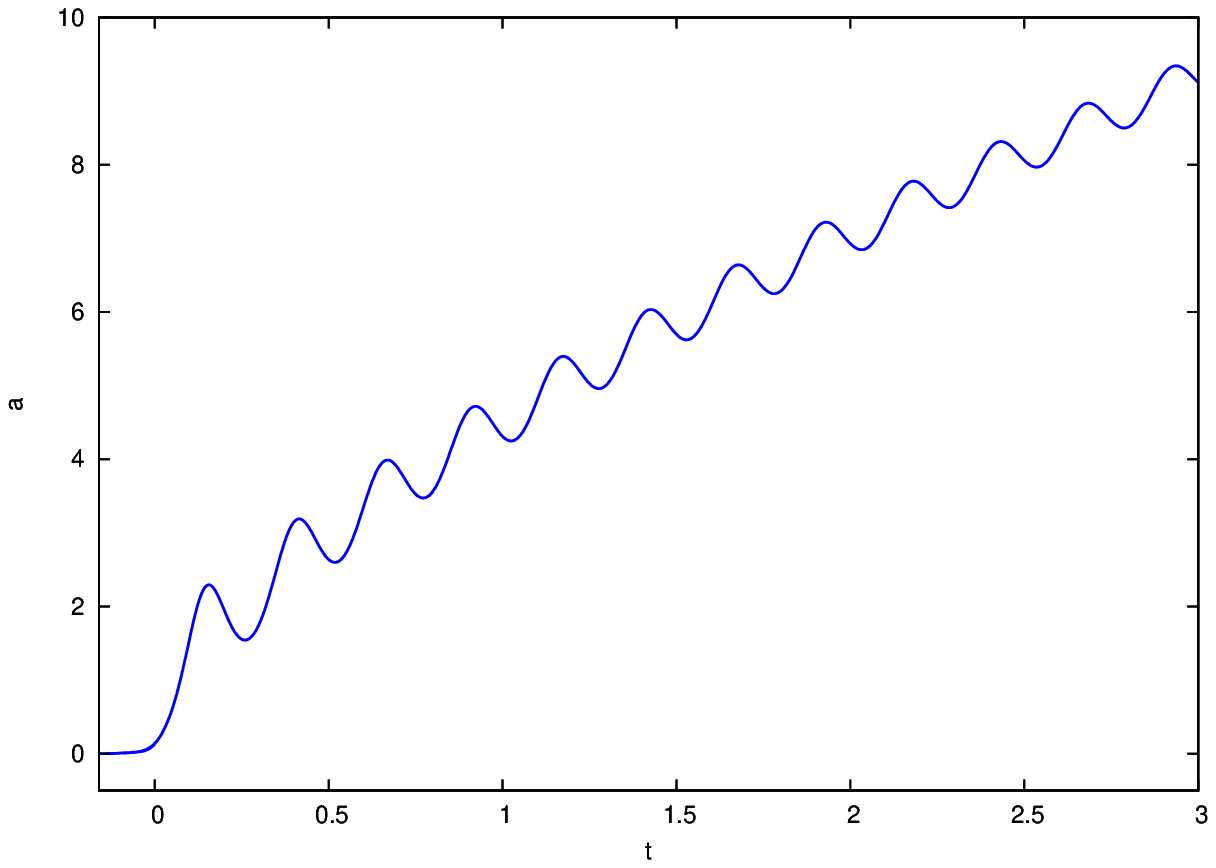}   
\end{array}
$
\caption{ 
{\bf Left  }: Temporal evolution of the coupling constant $g^2_s$. 
{\bf Right }: Evolution of the scale factor in the string frame.}
\label{Numeric3}
\end{figure}

\begin{figure}[ht!]
\centering
$\begin{array}{cc}   
\includegraphics[width=7cm,angle=0]{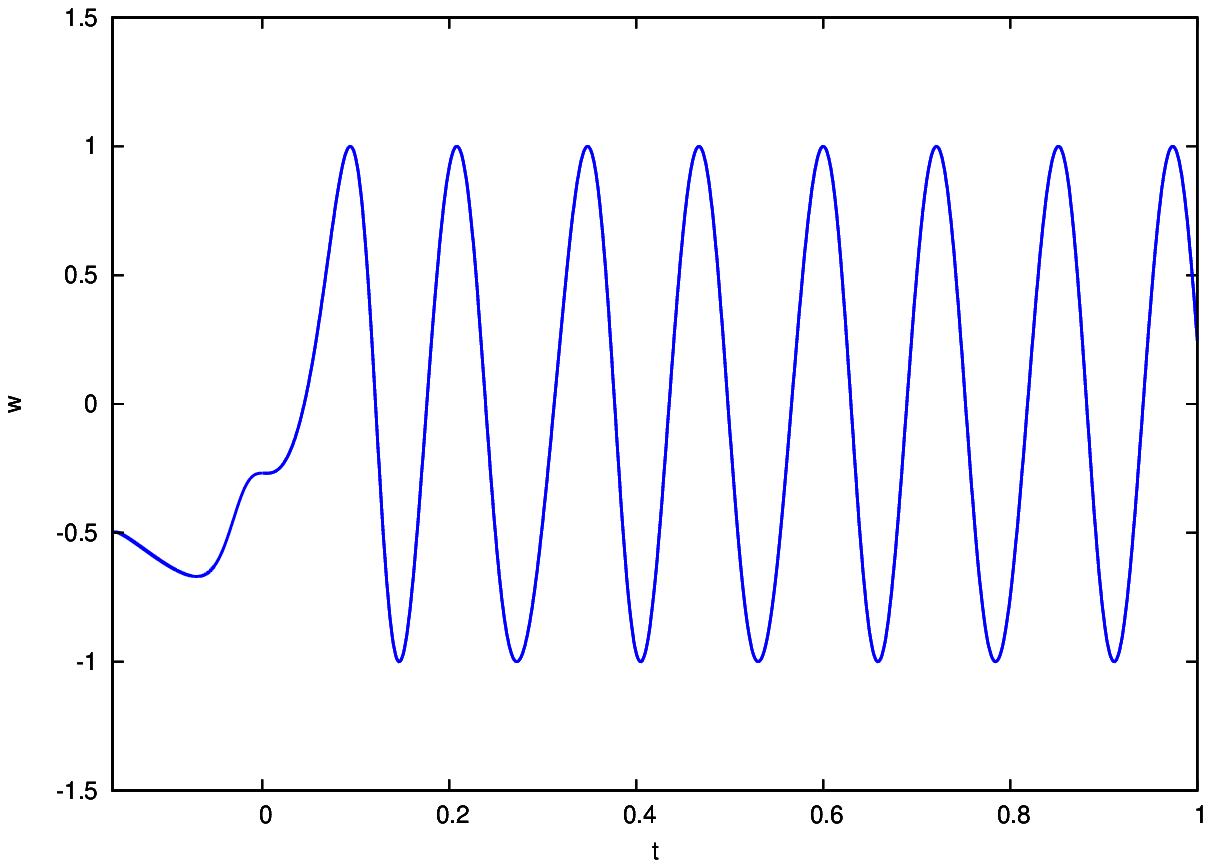}  &  \includegraphics[width=7cm,angle=0]{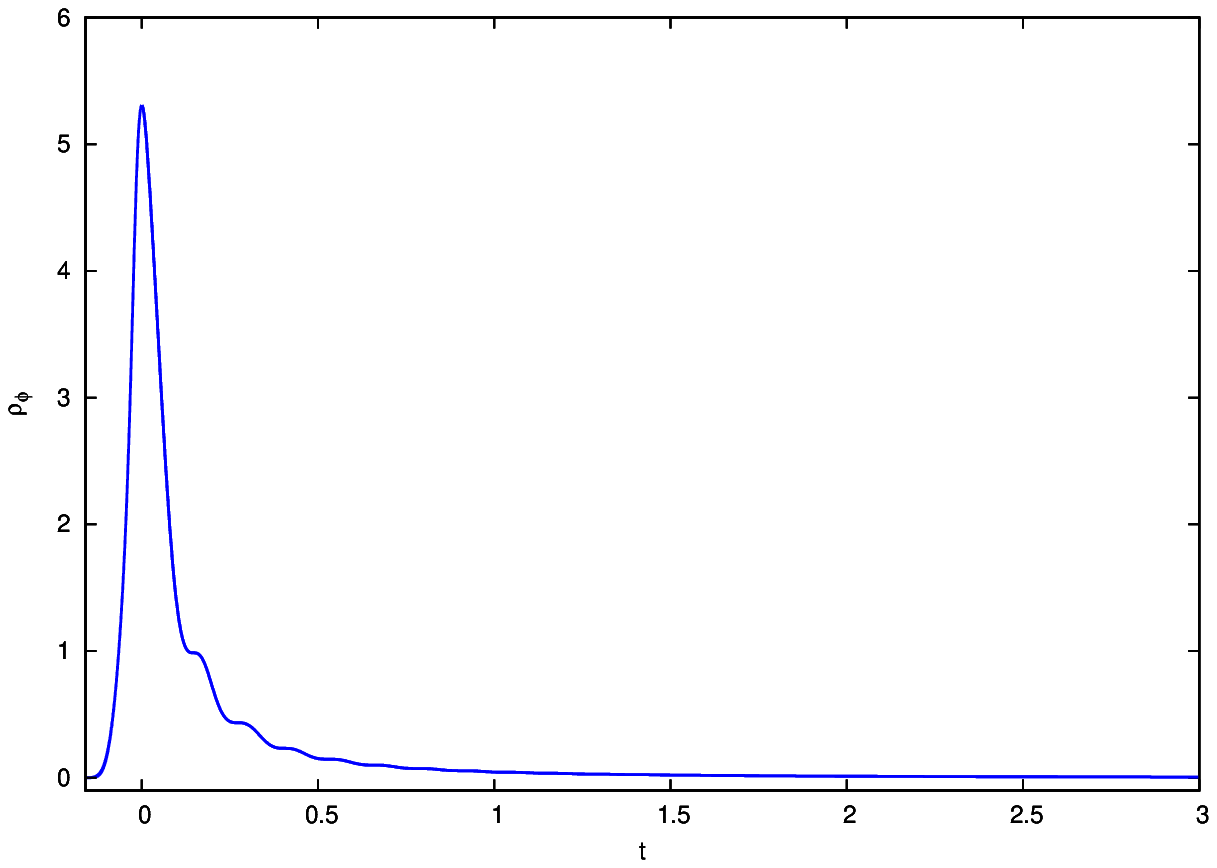}   
\end{array}
$
\caption{ 
{\bf Left  }: Evolution of the $w$ parameter in the equation of state $p=w\rho$. 
{\bf Right }: Evolution of the energy density of the dilaton field.}
\label{Numeric4}
\end{figure}

\begin{figure}[ht!]
\centering
$\begin{array}{cc}   
\includegraphics[width=7cm,angle=0]{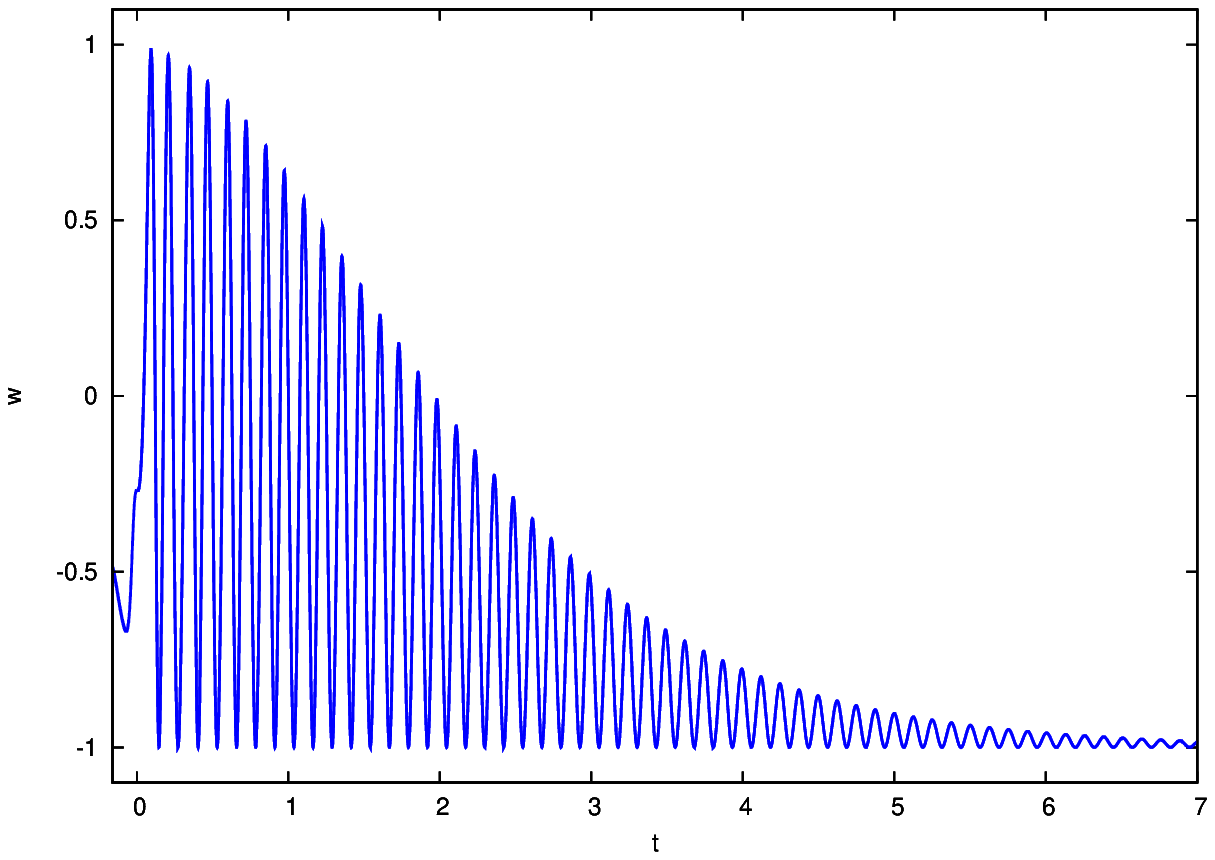}  &  \includegraphics[width=7cm,angle=0]{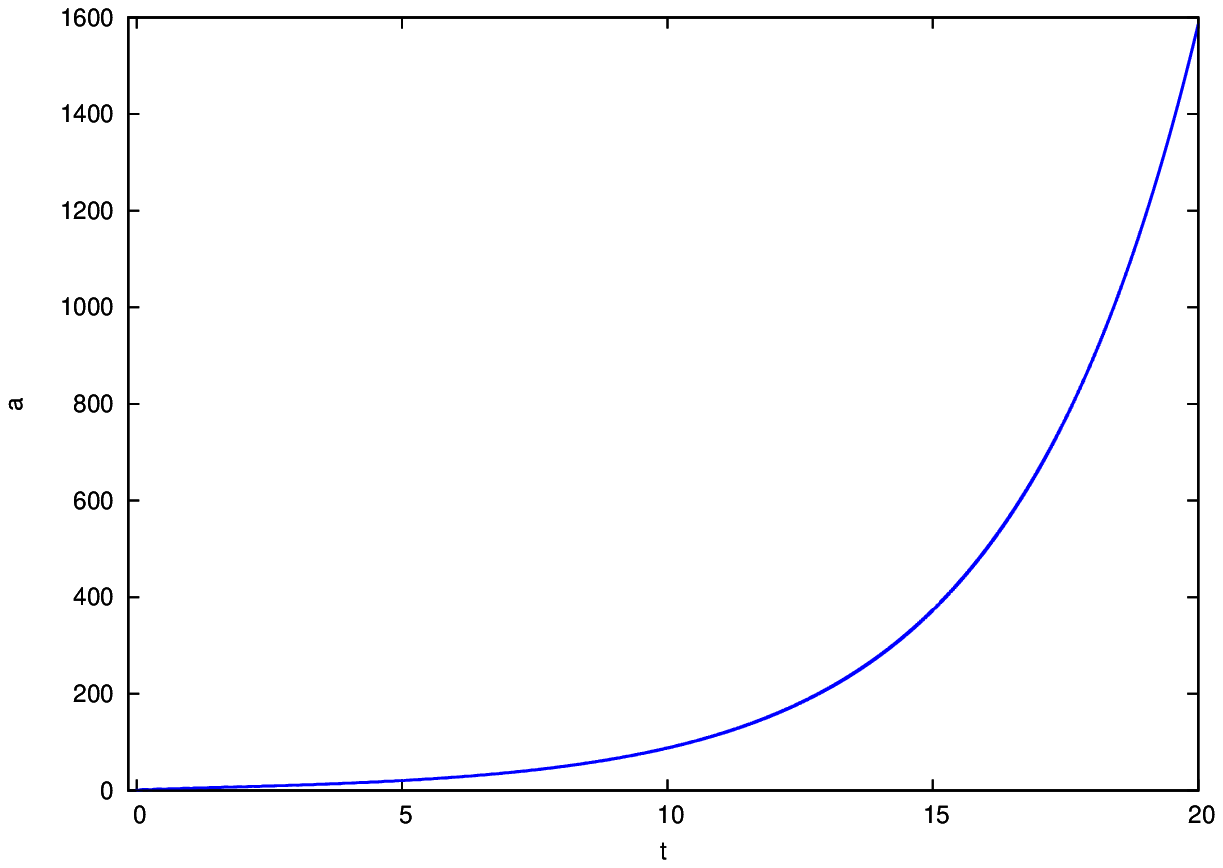}   
\end{array}
$
\caption{ 
{\bf Left  }: Evolution of the $w$ parameter in the equation of state $p=w\rho$ in the case $V_0=0.01/G^2$ . 
{\bf Right }: Corresponding evolution  of the scale factor in the string frame.}
\label{Numeric5}
\end{figure}

Now we choose $m = 25/\sqrt{G_4}$ and boundary conditions in the form 
\begin{eqnarray} 
\tilde{\phi}(\tilde{t}=0) &=& \tilde{\phi}_{\text{max}}, \\
\tilde{a}(\tilde{t}=0)    &=& 1.0,              \\
\rho(\tilde{t}=0)         &=& \rho_{\text{c}}, \\
t(\tilde{t}=0)            &=& 0.0 .
\end{eqnarray}
With such a setup we perform numerical integration of the equations
of motion. In Fig.~\ref{Numeric1} we show calculated evolution of the 
scale factor and the Hubble parameter in the Einstein frame. This behaviour
is qualitatively similar to the earlier studied case without the potential.
However here the symmetry $t \rightarrow -t $ is broken. 
In the left panel in Fig.~\ref{Numeric2} we show dependence between time in the 
string frame and time in the Einstein frame. We see  that for the positive times 
this dependence is one to one. This is the result of the dilaton stabilisation 
which gives $e^{\phi/2} \rightarrow 1$ in the transformation (\ref{timetrans}). 
This fact can be seen from the right panel in Fig.~\ref{Numeric2} where 
evolution of the dilaton field is shown.
Stabilisation of the dilaton field lead also to the saturation of the 
coupling constant $g^2_s$ as it was in the left panel in Fig.~\ref{Numeric3}.
In the right panel in Fig.~\ref{Numeric3} we show corresponding evolution 
of the scale factor in the string frame. In the small time scales 
this evolution undergoes oscillations resulting from the dilaton field damping 
oscillations.
In average scale factor behaves as for the universe dominated by dust
with the equation of state $p = w \rho$ where $\langle w\rangle = 0$.
The left panel in Fig.~\ref{Numeric4} show the evolution of the 
parameter $w$. In the right panel in Fig.~\ref{Numeric4} we show
evolution of the dilaton energy density $\rho$.
In the model considered above we have obtained expected saturation
of the coupling constant $g_s^2 \rightarrow$ const and classical 
post-Big Bang state with  $\langle w \rangle = 0$.  Now we will add
non-vanishing constant contribution $V_0\neq 0$ to the dilaton potential.
Such a modification is important only close to the local minimum point $\phi=0$ what 
correspond to the post-Big-Band phase. In the bottom of the potential well the dilaton field
behaves like dark energy with the equation of state $p=w\rho$ where 
\begin{equation}
w = \frac{\frac{\dot{\tilde{\phi}}^2}{2}- U(\tilde{\phi})}{\frac{\dot{\tilde{\phi}}^2}{2}+U(\tilde{\phi})} 
\rightarrow   \frac{-U(\tilde{\phi}_0)}{+U(\tilde{\phi}_0)} =-\frac{V_0}{V_0}= -1
\end{equation}
leading to the classical de Sitter phase. This scenario we study numerically in the case 
$V_0 = 0.01/G^2_4$. In the left panel in Fig. \ref{Numeric5} we show resulting 
behaviour of the parameter $w$, which is damped to the constant value $w=-1$. 
In the right panel in Fig. \ref{Numeric5} we show corresponding evolution of the 
scale factor in the string frame. 
The dilaton field can therefore behave like inflaton field or dark energy.

\section{Summary} \label{sec:summary}

In this paper we have analysed the low energy superstring theory with Loop Quantum Gravity effects 
as the tool to investigate the early Universe. This approach was initially proposed by de Rissi 
et al. \cite{De Risi:2007gp}. Based on this approach we had derived 
the cosmological model and solved it analytically. The important result which occurs 
is that an initial singularity state is avoided without need for the $t$-duality transformation.
In this scenario the universe starts its evolution from asymptotically free vacuum. Namely,
the state characterised by the vanishing coupling constant $g^2_s \rightarrow 0$, energy density 
$\rho \rightarrow 0$ and the scale factor $a \rightarrow 0$ for $t \rightarrow -\infty$.
This picture is the realisation of the ``asymptotic past triviality'' hypothesis. 
The universe starts its evolution from such a state and then evolves toward the 
non-singular Big Bang phase and subsequently to the current state.
However in this simple model the coupling constant monotonically grows during evolution. 
The effective evolution is asymptotically described by $a(t) \propto t^{1/\sqrt{3}}$ 
for $t \rightarrow \infty$. 

We have also proposed improvement of this model introducing potential for the 
dilaton field. In this case we also start from the initial asymptotically free 
vacuum. However in the further stages of the evolution, the dilaton field is 
constrained by the potential part $e^{\phi}$. After the non-singular Big Bang phase, 
characterised by the maximal and bounded energy density, the dilaton field ends 
its evolution in the bottom of potential well. This leads to the saturation of 
the coupling constant $g^2_s \rightarrow \text{const}$. The dilaton in this phase 
behaves like dark energy with the equation of state $p=-\rho$.  This leads to 
the emergence of the de Sitter phase which can explain present cosmological 
observations.

\begin{acknowledgments}
The authors are grateful to Abhay Ashtekar and Martin Bojowald for discussion. 
We would like to thank to Parampreet Singh and anonymous referee for 
important remarks.
This work has been supported by the Marie Curie Actions Transfer of Knowledge
project COCOS (contract MTKD-CT-2004-517186).
\end{acknowledgments}

\end{document}